\documentclass[twocolumn, prb, showpacs]{revtex4-1}

\usepackage{graphicx}
\usepackage{amsmath}
\usepackage{dcolumn}
\usepackage{hyperref}

\newcommand{\bh}{MgB$_{12}$H$_{12}$}
\newcommand{\MG}{Mg(BH$_4$)$_2$}

\begin{document}
%%%%%%%%%%%%%%%%%%%%%%%%%%%%%%%%%%%%%%%%%%%%%%%%%%%%%%%%%%%%%%%%%%%%%%%%

%%%%%%%%%%%%%%%%%%%%%%%%%%%%%%%%%%%%%%%%%%%%%%%%%%%%%%%%%%%%%%%%%%%%%%%%
\title{Tuning the hydrogen desorption\\ of \MG\ through Zn alloying}

\author{D. Harrison}
\affiliation{Department of Physics, Wake Forest University,
Winston-Salem, NC 27109, USA.}

\author{T. Thonhauser}
\email{thonhauser@wfu.edu}
\affiliation{Department of Physics, Wake Forest University,
Winston-Salem, NC 27109, USA.}

\date{\today}

\begin{abstract}
We study the effect of Zn alloying on the hydrogen desorption properties
of \MG\ using \emph{ab initio} simulations. In particular, we
investigate formation/reaction enthalpies/entropies for a number of
compounds and reactions at a wide range of temperatures and Zn
concentrations in Mg$_{1-x}$Zn$_x$(BH$_4$)$_2$. Our results show that
the thermodynamic stability of the resulting material can be
significantly lowered through Zn alloying. We find that e.g.\ the solid
solution Mg$_{2/3}$Zn$_{1/3}$(BH$_4$)$_2$ has a reaction enthalpy for
the complete hydrogen desorption of only 25.3~kJ/mol~H$_2$---a lowering
of 15~kJ/mol~H$_2$ compared to the pure phase and a corresponding
lowering in critical temperature of 123~K. In addition, we find that the
enthalpy of mixing is rather small and show that the decrease in
reaction enthalpy with Zn concentration is approximately linear.
\end{abstract}

\pacs{63.20.dk, 65.40.-b, 61.50.Ah}
% 63.20.dk      First-principles theory
% 65.40.-b      Thermal properties of crystalline solids
% 61.50.Ah      Theory of crystal structure, crystal symmetry;
%               calculations and modeling
% 88.30.rd      Inorganic metal hydrides
\maketitle
%%%%%%%%%%%%%%%%%%%%%%%%%%%%%%%%%%%%%%%%%%%%%%%%%%%%%%%%%%%%%%%%%%%%%%%%

%%%%%%%%%%%%%%%%%%%%%%%%%%%%%%%%%%%%%%%%%%%%%%%%%%%%%%%%%%%%%%%%%%%%%%%e
\section{Introduction}\label{sec:introduction}
%%%%%%%%%%%%%%%%%%%%%%%%%%%%%%%%%%%%%%%%%%%%%%%%%%%%%%%%%%%%%%%%%%%%%%%%

Metal borohydrides are amongst the most promising hydrogen storage
materials, \cite{Graetz_2009:new_approaches,
Ronnebro_2011:development_group, Li_2011:recent_progress,
Rude_2011:tailoring_properties, Jain_2010:novel_hydrogen} as they have
some of the highest storage densities. Unfortunately, the hydrogen
desorption temperature for the most attractive borohydrides is too high
for on-board storage, where it should be below
85$^\circ$C.\cite{DOE_Targets_Onboard_2009, Yang_2010:high_capacity}
Thus, borohydrides have been studied intensely in order to lower their
desorption temperature. Destabilizing via reactions with other hydrides
has been suggested, \cite{Vajo_2005:reversible_storage,
Alapati_2007:using_first, Alapati_2008:large-scale_screening,
Li_2008:dehydriding_rehydriding} as well as simple
doping,\cite{Nickels_2008:tuning_decomposition,
Hoang_2009:hydrogen-related_defects} cation
substitution,\cite{Setten_2007:ab_initio} anion
substitution,\cite{Brinks_2008:adjustment_stability} or adding
catalysts\cite{Li_2007:effects_ball}---but unfortunately, the results
still fall short of the required DOE
targets.\cite{DOE_Targets_Onboard_2009}

The borohydride \MG\ is of particular interest; it has substantial
storage density of 14.9~mass\% and its decomposition has been shown to
be fully reversible under certain
conditions,\cite{Severa_2010:direct_hydrogenation} but it completes its
first major intermediate step around 570~K and doesn't fully desorb
until it reaches temperatures around
820~K.\cite{Li_2008:dehydriding_rehydriding,
Soloveichik_2009:magnesium_borohydride} If we can lower its desorption
temperature, it might be one of the few materials to satisfy the DOE
targets. The desorption temperature is determined by the thermodynamics
and kinetics of the desorption reaction. Although it is generally
believed that kinetics is the ``culprit'' in the case of
\MG,\cite{Ozolins_2008:first-principles_prediction} it is still
desirable to first optimize the thermodynamics before addressing the
kinetic barrier.\cite{Alapati_2008:large-scale_screening} In the present
manuscript, we investigate the thermodynamics of the desorption of \MG\
and the dramatic effect Zn alloying has on it.

We have recently become aware of some very nice work also studying Mg/Zn
borohydride solid solutions, and will use this opportunity to compare
our results with theirs and point out similarities and
differences.\cite{Albanese_2013:theoretical_experimental} Further
simulations have been performed studying the desorption of
\MG,\cite{Zhang_2012:theoretical_prediction,
Setten_2008:density_functional} but without accounting for van der Waals
contributions, known to be necessary to achieve the correct energetic
ordering among polymorphs of
\MG;\cite{Huan_2013:thermodynamic_stability, Bil_2011:van_waals} we will
thus compare our results of the desorption pathway with other works to
ascertain the effect of van der Waals interactions. We also argue that
given the minimum practical delivery pressure from storage system of 3
bar, the overall desorption reaction is, in fact, outside the ideal
thermodynamic window of $-40$$^\circ$C to
+85$^\circ$C,\cite{DOE_Targets_Onboard_2009} but can be brought there by
alloying with Zn. We also find a linear relationship between the overall
hydrogen desorption enthalpy and Zn concentration in \MG.

Borohydrides are complex hydrides, in which tetrahedral anion units,
such as [BH$_4$]$^-$ and [AlH$_4$]$^-$, are bound to more
electropositive cations, such as Li, Na, K, Mg, and Ca. The ionic
bonding and charge transfer between the cations and the anion units is a
key feature of the stability of these
hydrides.\cite{Nakamori_2006:correlation_between} Targeting this
feature, the desorption temperature of Mg$_2$NiH$_4$ was successfully
lowered by destabilizing it through cation
substitution.\cite{Setten_2007:ab_initio} The same can also be achieved
by altering the anion complex, as demonstrated with fluorine
substitution in the hydrogen sublattice of
Na$_3$AlH$_6$.\cite{Brinks_2008:adjustment_stability,
Graetz_2009:new_approaches} More generally, an extensive experimental
study revealed an almost linear correlation between the desorption
temperature $T_d$ in K and the Pauling electronegativity $\chi_P$ as
well as the hydrogen desorption reaction enthalpy $\Delta H_r$ in
kJ/mol~H$_2$:\cite{T_d_fit}
\begin{eqnarray}
T_d &=& 1234 - 517.3\;\chi_P\;,\label{equ:fit_chi}\\
T_d &=& 423.4 + 8.34\;\Delta H_r\;\label{equ:fit_H}.
\end{eqnarray}
The study included data for $\mathcal{M}$(BH$_4$)$_n$ with $\mathcal{M}$
= Li, Na, K, Mg, Zn, Sc, Zr, and
Hf,\cite{Nakamori_2006:correlation_between} and was later extended by
Ca, Ti, V, Cr, and Mn.\cite{Nakamori_2007:thermodynamical_stabilities}
Zn(BH$_4$)$_2$ stands out in that it has the lowest temperature for full
decomposition of only 410~K.\cite{Jeon_2006:mechanochemical_synthesis}
Zn(BH$_4$)$_2$ itself is thermodynamically unstable at room temperature
and produces diborane gases in its decomposition reaction---it is thus
not directly interesting for hydrogen storage. But, as we will argue
below, Zn is an ideal alloyant for the \MG\ system, forming
Mg$_{1-x}$Zn$_x$(BH$_4$)$_2$, with remarkable effects. Zn alloying of
\MG\ is supported by the following facts: (i) Zn alloying has been found
experimentally to significantly lower the decomposition
temperature;\cite{Albanese_2013:theoretical_experimental,
Kalantzopoulos_2014:hydrogen_storage} (ii) Zn alloying lowers the
decomposition temperature of other
borohydrides;\cite{Ravnsbaek_2009:series_mixed-metal} (iii) Zn is known
to form a borohydride with the same stoichiometry as \MG\ and their
structures are essentially the same
\cite{Nakamori_2006:correlation_between} (the ionic radii of Zn and Mg
of 0.88 and 0.86 \AA\ are virtually identical); (iv) alloying in other
borohydrides, such as Li$_{1-x}$Cu$_x$BH$_4$, shows a desorption
temperature between the two
constituents;\cite{Miwa_2005:first-principles_study,
Nickels_2008:tuning_decomposition} (v) Zn is known to not form hydrides
or borides like other
borohydrides,\cite{Nakamori_2006:correlation_between} simplifying the
hydrogen desorption reaction significantly; and finally (vi) Zn is an
abundant major industrial metal.

%%%%%%%%%%%%%%%%%%%%%%%%%%%%%%%%%%%%%%%%%%%%%%%%%%%%%%%%%%%%%%%%%%%%%%%%
\section{Computational Details}
%%%%%%%%%%%%%%%%%%%%%%%%%%%%%%%%%%%%%%%%%%%%%%%%%%%%%%%%%%%%%%%%%%%%%%%%

In order to study the thermodynamics of \MG\ and the effect of Zn
alloying, we need the enthalpies and entropies for all materials
suspected in the decomposition of \MG. To this end, we performed {\it ab
initio} simulations at the DFT level, as
implemented in \textsc{Vasp},\cite{Kresse_1996:efficient_iterative,
Kresse_1999:ultrasoft_pseudopotentials} to calculate the ground-state
energies and phonon densities of states. We used PAW pseudopotentials
with a 650 eV kinetic energy cutoff and an energy convergence criterion
of $10^{-7}$~eV. We used $k$-point meshes giving convergence to within
1~meV/atom, with the exception of the metallic compounds and elemental
boron, which were converged to within 3~meV/atom. This yielded a
$k$-point mesh of e.g.\ $5\times5\times4$ for small unit cells like
MgB$_2$ and a mesh of $1\times1\times1$ for large unit cells like \MG.
Structures were relaxed with respect to unit-cell parameters and atom
positions until all forces were less than 0.1~meV/\AA. This way, we
found the lowest-energy structure of boron to be the 106 atom
$\beta$-rhombohedral structure suggested by van Setten et
al.\cite{Setten_2007:thermodynamic_stability} Phonons were calculated with the
symmetry-reduced finite-displacement method with displacements of
0.015~\AA. Supercells were created such that they were the same
dimensions as the $k$-point mesh used for the original unit cell. The
exceptions were the metals Mg and Zn, whose phonons were calculated with
a $5\times5\times2$ supercell and a $3\times3\times4$ $k$-point mesh.

While the ground-state structure of \MG\ is experimentally
known to be of P6$_1$22 symmetry with 30 formula units per unit cell,
theoretical studies with a variety of exchange-correlation (XC)
functionals find numerous other structures, all disagreeing with
experiment.\cite{Bil_2011:van_waals} The reason for this is that
\MG---similar to other borohydrides
\cite{Lodziana_2010:multivalent_metal}---exhibits a small but important
contribution from van der Waals interactions, on the order of 0.1~eV per
BH$_4$ unit. Including those contributions via the XC functional
vdW-DF\cite{Thonhauser_2007:van_waals,
Langreth_2009:density_functional}---with the (semi)local XC as
originally defined in Ref.~[\onlinecite{Dion_2004:van_waals}]---we were
the first to find the correct ground-state structure in agreement with
experiment.\cite{Bil_2011:van_waals} We further found that the closely
related P3$_1$12 structure is less than 1~meV per atom higher in energy.
This structure has only 9 formula units per unit cell and the local
coordination is almost identical to the P6$_1$22 structure, resulting in
almost identical phonon densities of states. Thus, in the present study we are
using vdW-DF and the numerically feasible P3$_1$12 structure for all our
simulations. This is further justified by comparing one of our reaction
enthalpies to the results of Albanese et al.\ in
Ref.~[\onlinecite{Albanese_2013:theoretical_experimental}], who in fact
used the large P6$_1$22 structure and obtained almost identical
results---see details below.  As our structure has 9 formula units per
unit cell, we report results for Zn concentrations in steps of 1/9.
Alloying was done by randomly replacing Mg atoms with Zn, with
variations---due to which atoms are being replaced---being on the order
of only fractions of a kJ/mol at the most. Our results for the enthalpy
of mixing compare well with Albanese et al., who found similarly small
values.\cite{Albanese_2013:theoretical_experimental}

%%%%%%%%%%%%%%%%%%%%%%%%%%%%%%%%%%%%%%%%%%%%%%%%%%%%%%%%%%%%%%%%%%%%%%%%
\section{Results}
%%%%%%%%%%%%%%%%%%%%%%%%%%%%%%%%%%%%%%%%%%%%%%%%%%%%%%%%%%%%%%%%%%%%%%%%

\begin{table*}
\caption{\label{tab:bader_analysis}Bader charges (in units of $e$) for
the metal site $\mathcal{M}$ and the BH$_4$ units as a function of
concentration $x$ in Mg$_{1-x}$Zn$_x$(BH$_4$)$_2$. Given are also the
Pauling electronegativity $\chi_P$ and the estimated desorption
temperature $T_d$ (in K) from Eq.~(\ref{equ:fit_chi}). The reported
charges and electronegativities are values averaged over all 9 formula
units in the unit cell.}
\begin{tabular*}{\textwidth}{@{}l@{\extracolsep{\fill}}rrrrrrrrrr@{}}\hline\hline
$x$           &    0    &    1/9  & 2/9     & 3/9     & 4/9     & 5/9     & 6/9     & 7/9     & 8/9     &   1   \\\hline
$\mathcal{M}$ &   +1.72 &   +1.65 & +1.59   & +1.53   & +1.46   & +1.40   & +1.33   & +1.26   & +1.20   &   +1.13\\
BH$_4$        & $-$0.86 & $-$0.83 & $-$0.79 & $-$0.76 & $-$0.73 & $-$0.70 & $-$0.67 & $-$0.63 & $-$0.60 & $-$0.56\\\hline
$\chi_P$      &    1.20 &    1.24 & 1.29    & 1.33    & 1.38    & 1.42    & 1.47    & 1.51    & 1.56    &    1.60\\
$T_d$         &    613  &    593  & 567     & 546     & 520     & 499     & 474     & 453     & 427     &    406 \\\hline\hline
\end{tabular*}
\end{table*}

\begin{figure}
\includegraphics[width=\columnwidth]{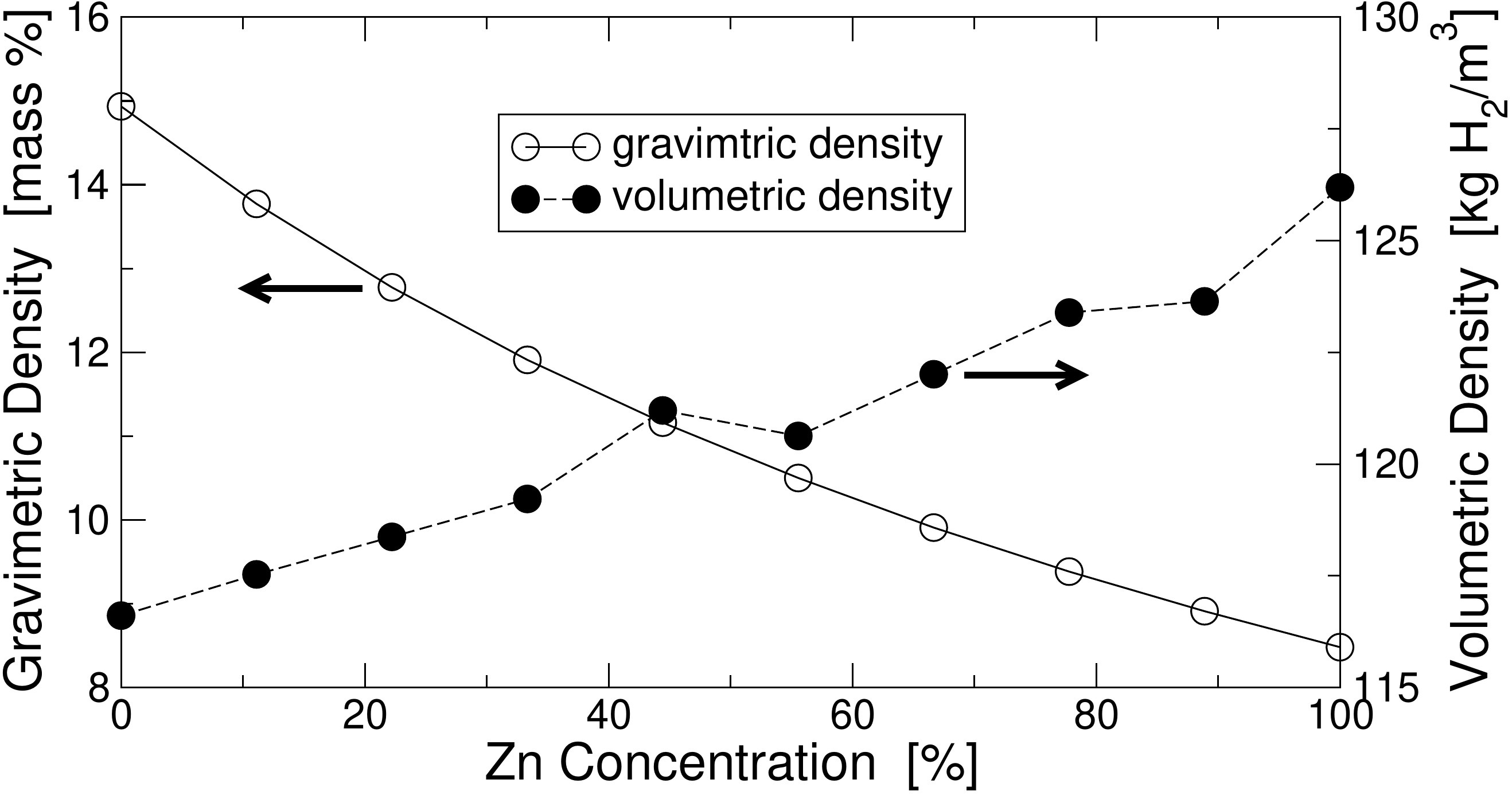}
\caption{\label{fig:storage_density}H$_2$ gravimetric storage density in
mass\% (to be read off the left axis) and H$_2$ volumetric storage
density in kg~H$_2$/m$^3$ (right axis) for different Zn concentrations.}
\end{figure}

%%%%%%%%%%%%%%%%%%%%%%%%%%%%%%%%%%%%%%%%%%%%%%%%%%%%%%%%%%%%%%%%%%%%%%%%
\subsection{Storage Densities}
%%%%%%%%%%%%%%%%%%%%%%%%%%%%%%%%%%%%%%%%%%%%%%%%%%%%%%%%%%%%%%%%%%%%%%%%

The gravimetric storage density of \MG\ will decrease with increasing Zn
content, as can be seen in Fig.~\ref{fig:storage_density}. But, as we
shall see below, only modest alloying is necessary to achieve the
required effect, and the loss in gravimetric density is reasonable. On
the other hand, the volumetric density increases with increasing Zn
content. The ``bump'' or deviation from linear behavior, observed in
Fig.~\ref{fig:storage_density} around 50\% Zn concentration, is also
seen in the work of Albanese et
al.,\cite{Albanese_2013:theoretical_experimental} and may be indicative
of the formation of a superstructure.

%%%%%%%%%%%%%%%%%%%%%%%%%%%%%%%%%%%%%%%%%%%%%%%%%%%%%%%%%%%%%%%%%%%%%%%%
\subsection{Structural Stability and Ionic Character}
%%%%%%%%%%%%%%%%%%%%%%%%%%%%%%%%%%%%%%%%%%%%%%%%%%%%%%%%%%%%%%%%%%%%%%%%

We begin by analyzing the structural stability in terms of the ionic
character as a function of Zn concentration. As mentioned above, the
ionic bonding is a key feature of the stability of borohydrides. In
Table~\ref{tab:bader_analysis} we quantify this picture and present a
Bader charge analysis as a function of Zn concentration, using the fast
implementation proposed by Henkelman et
al.\cite{Henkelman_2006:fast_robust} Note that the Bader analysis,
similarly to the Mulliken analysis, is an intuitive (but not unique) way
of partitioning the electron charge.  Interestingly, even for the pure
\MG\ structure, the ionic character significantly deviates from the
nominal values of +2 and $-1$. As expected, the ionic character
diminishes as a function of Zn concentration due to the higher
electronegativity of Zn, approximately 0.06~$e$ per 10\% Zn. A more
detailed analysis of the charge density reveals the following: While the
direct effect of Zn is localized, hydrogens further away from Zn get a
small compensating charge if they are within a tetrahedra next to a Zn.
Through this mechanism, even low levels of alloying influence almost the
entire structure. In Table~\ref{tab:bader_analysis} we also show the
Pauling electronegativity $\chi_P$ as a function of Zn concentration and
the corresponding estimated desorption temperatures $T_d$ according to
Eq.~(\ref{equ:fit_chi}). Purely based on this experimentally found
connection for borohydrides, we can already estimate that modest Zn
alloying might reduce the desorption temperature on the order of 100~K.
In the remainder of this paper, we will quantify this empirical
estimate.

%%%%%%%%%%%%%%%%%%%%%%%%%%%%%%%%%%%%%%%%%%%%%%%%%%%%%%%%%%%%%%%%%%%%%%%%
\subsection{Thermodynamics of the Hydrogen Desorption at 1 Bar Hydrogen
Pressure}
%%%%%%%%%%%%%%%%%%%%%%%%%%%%%%%%%%%%%%%%%%%%%%%%%%%%%%%%%%%%%%%%%%%%%%%%

We now move to the main point of this paper, i.e.\ the thermodynamics of
the hydrogen desorption and the effect of Zn alloying.  To this end, we
calculate the temperature dependent vibrational contribution to the
enthalpy and entropy as
\begingroup\small
\begin{eqnarray}
H_{\rm vib}&=&\int_0^\infty d\omega
   \bigg(\frac{1}{2}+\frac{1}{\exp[\hbar\omega/kT]-1}\bigg)
   g(\omega)\hbar\omega\;,\label{eq:enthalpy_phonon}\\[1ex]
S_{\rm vib} &=& \int_0^\infty d\omega
   \bigg(\frac{\hbar\omega}{2T}\coth\frac{\hbar\omega}{2kT}-k\ln\Big[2\sinh\frac{\hbar\omega}{2kT}\Big]\bigg)
   g(\omega)\;,\mspace{30mu}\label{eq:entropy_phonon}
\end{eqnarray}
\endgroup
where $\omega$ is the vibrational frequency, $g(\omega)$ is the phonon
density of states, $T$ is the temperature, and $k$ is Boltzmann's
constant. The enthalpy then is the sum of the DFT ground-state energy
and this vibrational contribution. Formation enthalpies, in turn, are
calculated as differences in enthalpies of the material and its
constituent elements (thus, elements in their natural state have formation enthalpies of
0). From this, we calculate enthalpies and entropies of reaction, using
formation enthalpies and absolute entropies of all materials
(tabulated in the appendix). For reactions involving Zn
alloying (Reactions~2, 3, and 13 in Table~\ref{tab:react_values}), the
entropy of mixing was calculated according to
\begin{equation}
S_{\rm mix}= -k_B\; [c\ln{c}+(1-c)\ln{(1-c)}]\;,
\end{equation}
where $c$ is the concentration of Zn; this entropy of mixing was added
to the entropy of reaction calculated from vibrational frequencies,
resulting in e.g.\ a decrease in reaction entropy of
$\sim5$~J/K/mol~H$_2$ and a corresponding increase in the critical
temperature of $\sim10$~K.  Note that all structures we found are true
local minima and none of the density of states show imaginary
frequencies.

Following the approach of van Setten et
al.,\cite{Setten_2008:density_functional} the temperature-dependent
thermodynamic values of H$_2$ were obtained using experimental
data.\cite{Hemmes_1986:thermodynamic_properties} In particular, we used
$H_{\rm{H_2\,gas}}(T)=E_{\rm{H_2}}+E^{\rm{ZPE}}_{\rm{H_2}}+H^{\rm{exp}}_{\rm{H_2\,gas}}(T)$,
where the electronic energy $E_{\rm{H_2}}$ and zero-point energy
$E^{\rm{ZPE}}_{\rm{H_2}}$ were calculated using DFT and the last term
was taken from experiment.\cite{Hemmes_1986:thermodynamic_properties}
Specifically, data was taken from H$_2$ at 1~bar for increments of 100~K
with values in between being linearly interpolated. Note that the
enthalpy of H$_2$ changes very little over moderate pressure changes
(e.g. the change in enthalpy going from 1 to 100~bar at 300~K is only
0.114~kJ/mol), meaning our formation enthalpies should be useful even
when looking at high pressures. The entropy of hydrogen gas used to
calculate the entropies of reaction in Table~\ref{tab:react_values}
and the reactions at 3~bar discussed later on were likewise taken from
the same experimental data.\cite{Hemmes_1986:thermodynamic_properties}
For reference, the entropies of H$_2$ at 300~K for 1 and 3~bar are
130.77 and 121.63~J/K/mol~H$_2$, respectively. Note that the entropy
of hydrogen gas for other pressures can be accurately estimated by the
equation $S_{\rm{H_2}}=-R\ln{p}+S_0$, where $p$ is the pressure in bar
and $S_0$ is the entropy at 1~bar.

\begin{table*}
\caption{\label{tab:react_values}Reaction enthalpies $\Delta H_r$ in
kJ/mol~H$_2$ and entropies $\Delta S_r$ in J/K/mol~H$_2$ at 300 K for
several \MG\ desorption reactions. The critical temperature $T_c$
predicted from the van't Hoff equation $\ln{p}=-\Delta H/RT+\Delta S/R$
for 1 bar H$_2$ pressure is given in K. We give a rough estimate of the
kinetic barrier in K as the difference $T_{\textrm{barrier}}=T_d-T_c$,
where $T_d$ is the approximate experimental desorption temperature.}
\begin{tabular*}{\textwidth}{@{}lr@{\quad$\longrightarrow$\quad}l@{\extracolsep{\fill}}rrrrr@{}}\hline\hline
No.& Reactants                                             & Products                                                                               & $\Delta H_r^{\rm 300K}$ & $\Delta S_r^{\rm 300K}$ & $T_c$ & $T_d$ & $T_{\textrm{barrier}}$\\\hline
1  & \MG\                                                  & $\textrm{MgB}_2+4\textrm{H}_2$                                                         & 40.3            & 112.15          & 360 & 820$^a$ & 460\\
2  & Mg$_{2/3}$Zn$_{1/3}$(BH$_4$)$_2$                      & $\frac{2}{3}\textrm{MgB}_2+\frac{1}{3}\textrm{Zn}+\frac{2}{3}\textrm{B}+4\textrm{H}_2$ & 25.3            & 106.72          & 237   \\
3  & Mg$_{1/3}$Zn$_{2/3}$(BH$_4$)$_2$                      & $\frac{1}{3}\textrm{MgB}_2+\frac{2}{3}\textrm{Zn}+\frac{4}{3}\textrm{B}+4\textrm{H}_2$ & 10.4            & 106.57          & 98   \\
4  & Zn(BH$_4$)$_2$                                        & $\textrm{Zn}+2\textrm{B}+4\textrm{H}_2$                                                & $-3.92$         & 112.57          & $-35$  & 410$^b$ & 445\\
\hline\noalign{\vskip 0.3mm}
5  & \MG\                                                  & $\frac{1}{6}\textrm{\bh}+\frac{5}{6}\textrm{MgH}_2+\frac{13}{6}\textrm{H}_2$           & 24.8            & 104.71   & 237 & 570$^c$ & 333   \\
6  & MgH$_2$                                               & $\textrm{Mg}+\textrm{H}_2$                                                             & 78.7            & 132.99   & 592 & 640$^d$ & 48  \\
7  & $\frac{1}{6}\textrm{\bh}$                             & $\frac{1}{6}\textrm{Mg}+2\textrm{B}+\textrm{H}_2$                                      & 85.4            & 119.30   & 716 & 730$^d$ & 14 \\
\noalign{\vskip 0.3mm}\hline\noalign{\vskip 0.3mm}
8  & \MG\                                                  & $\frac{1}{6}\textrm{\bh}+\frac{5}{6}\textrm{Mg}+3\textrm{H}_2$                         & 39.8            & 112.56   & 353   \\
9  & $\frac{1}{6}\textrm{\bh}+\frac{5}{6}\textrm{Mg}$      & $\textrm{MgB}_2+\textrm{H}_2$                                                          & 42.0            & 110.91   & 379   \\
10 & $\frac{1}{6}\textrm{\bh}$                             & $\frac{1}{6}\textrm{MgH}_2+2\textrm{B}+\frac{5}{6}\textrm{H}_2$                        & 86.8            & 116.56   & 744   \\  
11 & $\frac{1}{6}\textrm{\bh}+\frac{5}{6}\textrm{MgH}_2$   & $\frac{1}{2}\textrm{MgB}_4+\frac{1}{2}\textrm{MgH}_2+\frac{4}{3}\textrm{H}_2$          & 63.5            & 121.13   & 525   \\
\noalign{\vskip 0.3mm}\hline\noalign{\vskip 0.3mm}    
12 & \MG\                                                  & $\textrm{MgH}_2+2\textrm{B}+3\textrm{H}_2$                                             & 42.0            & 108.00   & 389   \\
13 & Mg$_{4/5}$Zn$_{1/5}$(BH$_4$)$_2$                      & $\frac{4}{5}\textrm{MgH}_2+\frac{1}{5}\textrm{Zn}+2\textrm{B}+\frac{16}{5}\textrm{H}_2$& 30.5            & 104.98   & 291   \\\noalign{\vskip 0.3mm}\hline\hline
\end{tabular*}
\raggedright $^a$Ref.~[\onlinecite{Soloveichik_2009:magnesium_borohydride}]; $^b$Ref.~[\onlinecite{Jeon_2006:mechanochemical_synthesis}]; $^c$Ref.~[\onlinecite{Yan_2008:differential_scanning}]; $^d$Ref.~[\onlinecite{Li_2008:dehydriding_rehydriding}].
\end{table*}

The overall energy required for the hydrogen decomposition reaction to
start can be split into two pieces, i.e.\ the \emph{reaction enthalpy}
and an additional \emph{kinetic barrier}. We argue that the latter is
similar for the initial hydrogen release of many decomposition reactions
of borohydrides and focus first on the reaction enthalpies.
Table~\ref{tab:react_values} contains the most pertinent results from
this study; it lists reaction enthalpies for a number of possible
reactions, calculated from formation enthalpies for a range of
materials. Where a direct comparison with experiment is possible, we
generally find very good agreement. For example, experimental formation
enthalpies for MgH$_2$ and MgB$_2$ of $-75.3$ and $-41.2$~kJ/mol are in
excellent agreement with our calculated values of $-78.7$ and
$-43.4$~kJ/mol (see appendix A).\cite{Vajo_2004:altering_hydrogen,
Balducci_2005:thermodynamics_intermediate} Furthermore, our calculated
reaction enthalpies for Reactions 1 and 8 of
40.3~kJ/mol~H$_2$ and 39.8~kJ/mol H$_2$ are also in good agreement with
the value of $40\pm2$~kJ/mol~H$_2$ found by Yan et al.\ in
Ref.~[\onlinecite{Yan_2008:differential_scanning}].\footnote{The authors
of Ref.~[\onlinecite{Yan_2008:differential_scanning}] propose here the
decomposition of \bh\ before MgH$_2$, which we find to be unlikely as it
has a critical temperature of 716~K.  We suspect the actual reaction
studied was the one we proposed.} Previous theoretical studies have
calculated an overall reaction enthalpy of 38--39~kJ/mol~H$_2$ at room
temperature,\cite{Setten_2008:density_functional,
Zhang_2012:theoretical_prediction,
Ozolins_2008:first-principles_prediction,
Ozolins_2009:first-principles_prediction} which closely agrees with our
value of 40.3~kJ/mol~H$_2$. We attribute the difference to vdW-DF, which
stabilizes \MG\ with respect to its reaction products by accounting for
the long-range van der Waals forces. Comparison of our results for other
decomposition reactions with previous studies shows similar agreement,
with differences on the order of several kJ/mol~H$_2$ due to vdW-DF.
Although accounting for long-range van der Waals forces has been found
to be critical to finding the correct energetic ordering among
polymorphs in borohydrides,\cite{Huan_2013:thermodynamic_stability,
Bil_2011:van_waals} it seems to affect total reaction enthalpies by only
several kJ/mol~H$_2$ at the most.

\begin{figure}
\includegraphics[width=\columnwidth]{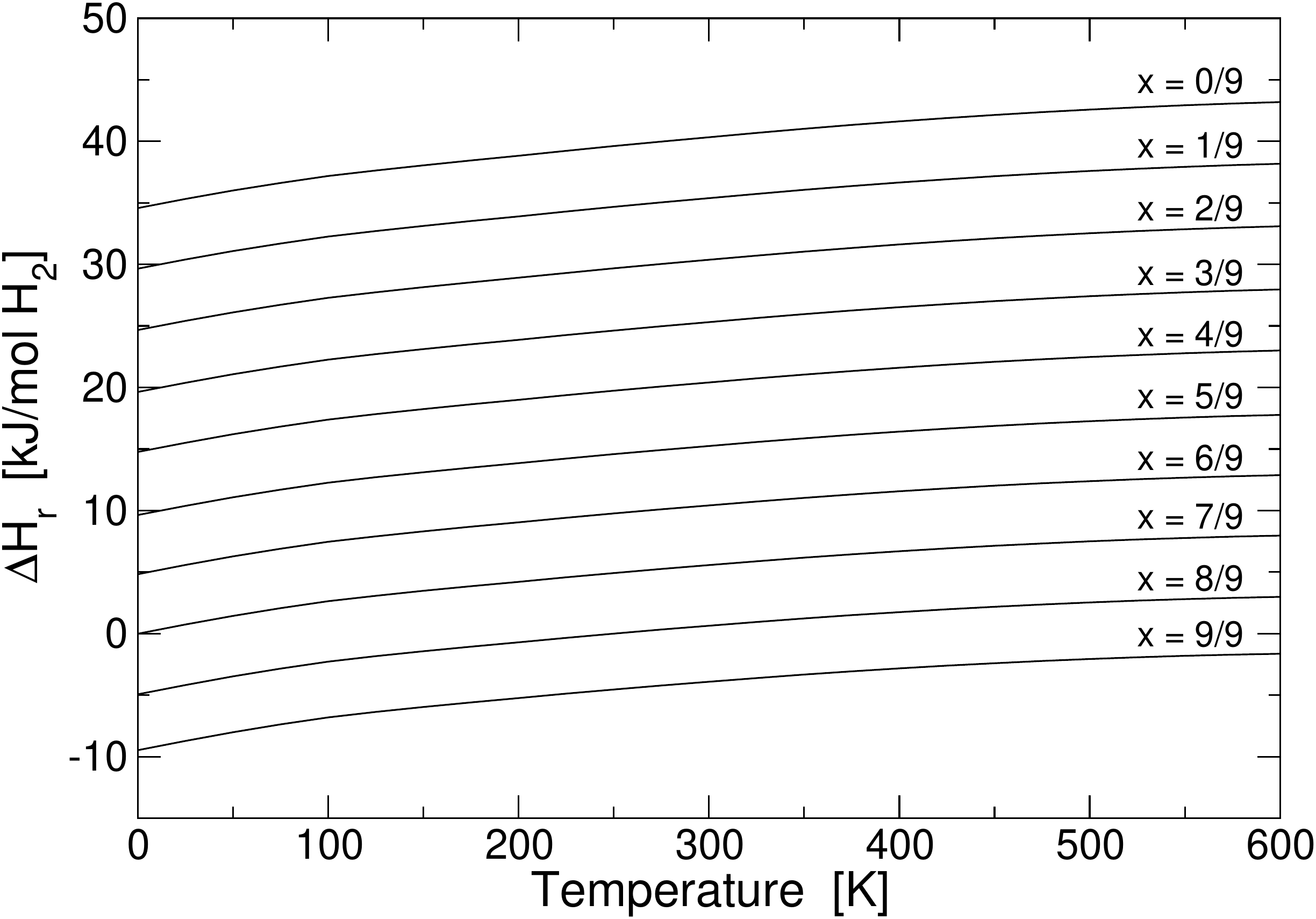}
\caption{\label{fig:reaction_enthalpies}Reactions enthalpies
$\Delta H_r$ can readily be calculated from the
tabulated data in the appendices. As an example, we plot here the
reaction enthalpy as a function of temperature and alloying for the
reaction $\textrm{Mg$_{1-x}$Zn$_x$(BH$_4$)$_2$} \rightarrow
\textrm{Mg$_{1-x}$B$_{2(1-x)}$} + \textrm{Zn$_x$} + 2\textrm{B$_x$} +
4\textrm{H$_2$}$. Note that the effect of Zn concentration on the
reaction enthalpy is almost perfectly linear.}
\end{figure}

Reactions 1 -- 4 of Table~\ref{tab:react_values} show the effects of Zn
alloying on the overall reaction enthalpy. We find the effect to be
linear and every 10\% Zn results in a further lowering of the desorption
enthalpy by approximately 4.5~kJ/mol~H$_2$, as can be seen in
Fig.~\ref{fig:reaction_enthalpies}. Zn
alloying thus provides a convenient way of tuning the desorption
enthalpy over a wide range. Note that the desorption enthalpy of
Reactions 2 and 5 are approximately the same, but the important
difference is that the alloyed phase releases the entire hydrogen, while
the latter only releases half of the available hydrogen. Furthermore,
note that Reactions 1 -- 4 are different from the decomposition modeled
in Ref.~[\onlinecite{Albanese_2013:theoretical_experimental}] in that we
model the entire hydrogen release, i.e.\ the outcome is MgB$_2$ +
4H$_2$, whereas they assume the intermediate MgH$_2$. For comparison, we
have modeled their reaction also (Reaction 13 in
Table~\ref{tab:react_values}), obtaining 30.5~kJ/mol~H$_2$ in excellent
agreement to their value of 30~kJ/mol~H$_2$---in fact, justifying
approximations that both our groups have made.  It is important to also
compare the two corresponding ``pure'' reactions, i.e.\ Reaction 1 and
12. The reaction enthalpy for the entire hydrogen release in Reaction 1
is, in fact, a little lower compared to Reaction 12, which led us to
pursue the pure reaction pathway.

Note that we did not list the effects of Zn alloying on intermediate
reactions because it was unknown whether Zn would remain in \bh\ or
phase separate, and in either case there was no experimental data on the
corresponding structures.  Reactions 5 -- 7 show suspected intermediate
steps, while Reactions 8 -- 13 show alternative reactions discussed
later; data for other possible reactions can be readily calculated using
extensive thermodynamic data tabulated in the appendix. 

Of course, it is also well known that there are more intermediates in
the decomposition of \MG\ involving Mg(B$_x$H$_y$)$_z$
complexes,\cite{Ozolins_2008:first-principles_prediction,
Newhouse_2010:reversibility_improved} especially in the formation and
decomposition of MgB$_{12}$H$_{12}$. While we did not study all of these
possible intermediates, our calculated thermodynamic values in the
appendix can give insight into some other favorable
pathways. Within the framework of pathways considered, our results show
that the most favorable pathway for hydrogen desorption that agrees with
experiment is:
\begin{align}
\textrm{Mg(BH$_4$)$_2$} &\rightarrow \frac{1}{6}\textrm{MgB$_{12}$H$_{12}$}\label{eq:reaction_mgb12h12}
   +\frac{5}{6}\textrm{MgH$_{2}$}
   + \frac{13}{6}\textrm{H}_2\\
&\rightarrow \frac{1}{6}\textrm{MgB$_{12}$H$_{12}$}
   \label{eq:reaction_mg}
   +\frac{5}{6}\textrm{Mg}
   + 3\textrm{H}_{2}\\
&\rightarrow \textrm{Mg}
   \label{eq:reaction_elements}
   +2\textrm{B}
   + 4\textrm{H}_{2}\\   
&\rightarrow \textrm{MgB$_{2}$} +
    4\textrm{H$_2$}\label{eq:reaction_mgb2}
\end{align}
Thermodynamically, as seen in Reaction 9, the most favorable pathway
after the formation of \bh\ is the formation of MgB$_2$, but from
experimental differential scanning calorimetry and chemical intuition,
it is likely that there is at least one intermediate step from \bh\ to
MgB$_2$. The theoretical total hydrogen loss for steps
\eqref{eq:reaction_mgb12h12}, \eqref{eq:reaction_mg}, and
\eqref{eq:reaction_mgb2} is 8.1\%, 11.2\%, and 14.9\%, respectively.
The formation of \bh, MgH$_2$, Mg, and MgB$_2$ are all well-known steps
confirmed by X-ray diffraction and solid state
NMR.\cite{Li_2008:dehydriding_rehydriding, Yan_2011:formation_process,
Soloveichik_2009:magnesium_borohydride, Yang_2011:decomposition_pathway}
They are also supported as thermodynamically favorable by a number of
theoretical studies.\cite{Zhang_2012:theoretical_prediction,
Zhang_2010:theoretical_prediction} Furthermore, the temperature
evolution of the reaction products also supports this reaction pathway:
MgH$_2$ appears first, then Mg without MgH$_2$, and finally MgB$_2$. 

Equation~\eqref{eq:reaction_mgb12h12} also corresponds to the mass\% of
hydrogen desorbed for certain temperatures held for long periods of
time. Yan et al.\ found that just over 8 mass\% of H$_2$ desorb after
1000~min at 573~K; this corresponds almost exactly to the expected value
of 8.1\%\ for
Eq.~\eqref{eq:reaction_mgb12h12}.\cite{Yan_2011:formation_process}  It
is also worth noting that it took over an hour before even 4 mass\% of
H$_2$ was desorbed, and the curves for lower temperatures never
equilibrated after 1000~min. Due to the sluggish kinetics of this step
of the reaction it is difficult to see the individual reaction steps in
a typical thermogravimetry curve; by the time only a fraction of \MG\
has undergone the first step of the reaction, the temperature has
increased so that the second step of the reaction has already begun.
However, we can see further evidence for the proposed reaction pathway
by the results of Matsunaga et al., who found that \MG\ could be
rehydrogenated from 11\%\ to 8\%\ at 623~K, which corresponds to the
rehydrogenation of Mg to MgH$_2$ (see Fig.~6 of
\nocite{Matsunaga_2008:hydrogen_storage}
Ref.~[\onlinecite{Matsunaga_2008:hydrogen_storage}]). 

We also find that the previously suggested decomposition of \bh\ to
MgH$_2$ and elements shown in Reaction~10 to be unfavorable as a second
step of the reaction, with a critical temperature of 744~K, and suggest
the decomposition of MgH$_2$ ($T_c=592$~K) as occurring before \bh.
After MgH$_2$ decomposes, we suggest the decomposition of \bh\, as our
critical temperature of 716~K coincides closely with the final large dip
observed in differential scanning calorimetry plots near 1 bar H$_2$
pressure.\cite{Yan_2008:differential_scanning} Finally, we find the
formation of MgB$_4$ to be plausible, as evidenced by the favorable
thermodynamics of Reaction~11.

Reaction~13 is the decomposition reaction modeled by Albanese et al.\ as
having the optimal decomposition
enthalpy;\cite{Albanese_2013:theoretical_experimental} our calculated
value of 30.5~kJ/mol~H$_2$ is in exact agreement with their result of
30~kJ/mol~H$_2$. Note that our enthalpy and entropy used for 20\%\
Zn-alloyed \MG\ were calculated from a linear interpolation of our
thermodynamic data; this is valid because the change in enthalpy and
entropy values with Zn concentration is nearly exactly linear (see the
appendix). It is interesting to see that, when we
interpolate our results for Reactions 1 -- 4 to a Zn concentration of
1/5 to match the concentration in Reaction 13, we find a reaction
enthalpy and entropy of 31.46~kJ/mol~H$_2$ and 112~J/K/mol~H$_2$,
leading to a critical temperature $T_c=281$~K, which is essentially the
same as for Reaction 13 in Table~\ref{tab:react_values}.  Although our
results for Reaction~13 agree with the work done by Albanese et al., our
work can be seen as an extension of the theoretical part of that study,
as we have actually calculated the full thermodynamic data---including
reaction entropies---instead of interpolating them. We have done so for
their proposed partial decomposition (Reaction 13) and in addition for a
large number of other reactions (Reactions 1 -- 11), including the full
hydrogen decomposition as well as others that can easily be deduced from
the tabulated thermodynamic data in our appendix.

%%%%%%%%%%%%%%%%%%%%%%%%%%%%%%%%%%%%%%%%%%%%%%%%%%%%%%%%%%%%%%%%%%%%%%%%
\subsection{Thermodynamics of the Hydrogen Desorption at 3 Bar Hydrogen
Pressure}
%%%%%%%%%%%%%%%%%%%%%%%%%%%%%%%%%%%%%%%%%%%%%%%%%%%%%%%%%%%%%%%%%%%%%%%%

From the results above it follows that kinetics is not the only problem
with \MG, as the reaction from \MG\ to MgB$_2$ has a calculated critical
temperature of 360~K (86$^{\circ}$C). Note that this critical
temperature was calculated at 1~bar~H$_2$ pressure, while the minimum
DOE target for delivery pressure is around 3~bar; the critical
temperature at 3 bar is 430~K (157$^{\circ}$C), well outside the optimal
thermodynamic window. Furthermore, while Zuttel et al.\ have argued that
the entropic contribution to metal hydride reactions is approximately
130~J/K/mol~H$_2$ for most simple metal-hydrogen
systems,\cite{Zuttel_2003:libh4_new} lower values can occur for complex
metal hydrides, such as 97~J/K/mol~H$_2$ for
$\textrm{LiBH}_4\rightarrow\textrm{LiH}+\textrm{B}+3/2\textrm{H}_2$.\cite{Alapati_2007:using_first}
It seems likely then that \MG\ also has a lower than normal entropy of
reaction, in agreement with our calculated values in
Table~\ref{tab:react_values}, given its similarity to LiBH$_4$.  Because
of this comparatively low value for the reaction entropy, even the
commonly cited value of 40~kJ/mol H$_2$ is too high for a practical PEM
fuel cell. In fact, using our calculated entropy value, for a minimum
delivery pressure of 3 bar and temperature of 80$^\circ$C, the maximum
desired enthalpy is 33~kJ/mol~H$_2$. Also note that while the estimated
desired reaction enthalpy for a hydrogen storage material is
20--50~kJ/mol~H$_2$ (for reactions with low to high entropic
contributions), the ideal range for efficiency is
20--30~kJ/mol~H$_2$.\cite{Yang_2010:high_capacity} All of this points
towards the conclusion that \MG\ needs to be destabilized with respect
to its reaction products in order to achieve the ideal thermodynamic
reaction window. E.g.\ in the case of 1/3 Zn alloying, we find a
reaction enthalpy of 25.3~kJ/mol~H$_2$ and a critical temperature of
270~K ($-3^{\circ}$C) at 3 bar, both in the ideal range for a PEM fuel
cell.  This is based on the suspected decomposition reaction
$\textrm{Mg$_{1-x}$Zn$_x$(BH$_4$)$_2$}\rightarrow\textrm{Mg$_{1-x}$B$_{2(1-x)}$}
+ \textrm{Zn$_x$}+2\textrm{B$_x$}+ 4\textrm{H$_2$}$. We find this to be
the most likely reaction as Zn(BH$_4$)$_2$ ``decomposes directly to
elemental Zn due to instabilities of Zn hydride and
boride,''\cite{Nakamori_2006:correlation_between} and we find Zn-alloyed
MgB$_2$ to be unstable and expect it to phase separate to MgB$_2$, Zn,
and B based on preliminary calculations. Of course, for high
concentrations of Zn, we expect the formation of diborane, as seen
experimentally by Albanese et al.\ and Kalantzopoulos et
al.,\cite{Albanese_2013:theoretical_experimental,
Kalantzopoulos_2014:hydrogen_storage} but for low concentrations of Zn
it is known that diborane is not formed.

%%%%%%%%%%%%%%%%%%%%%%%%%%%%%%%%%%%%%%%%%%%%%%%%%%%%%%%%%%%%%%%%%%%%%%%%
\subsection{Kinetics of the Hydrogen Desorption and overall Desorption
Temperature}
%%%%%%%%%%%%%%%%%%%%%%%%%%%%%%%%%%%%%%%%%%%%%%%%%%%%%%%%%%%%%%%%%%%%%%%%

We now switch to a discussion of the kinetic barrier. By comparing our
calculated critical temperatures $T_c$ with the known experimental
temperatures of hydrogen desorption
$T_d$,\cite{Li_2008:dehydriding_rehydriding, Yan_2011:formation_process,
Soloveichik_2009:magnesium_borohydride,
Jeon_2006:mechanochemical_synthesis} we can make a rough estimate of the
kinetic barrier, given in Table~\ref{tab:react_values}. Interestingly,
from Eq.~(\ref{equ:fit_H}) we see that---by definition---borohydrides
with $\Delta H_r = 0$ have a kinetic barrier of 423.4~K; using the same
argument but a slightly different fit (see footnote \citenum{T_d_fit}),
we find a kinetic barrier of 439.4~K. Borohydrides with $\Delta H_r
\approx 0$, such as Zn(BH$_4$)$_2$, thus must have a kinetic barrier of
the same order of magnitude, in very good agreement to our calculated
value of 445~K for Zn.

From Table~\ref{tab:react_values} we see that the kinetic barrier for
the full hydrogen release reaction is almost the same for \MG\ and
Zn(BH$_4$)$_2$, which in turn means that differences in calculated
critical temperatures in Table~\ref{tab:react_values} approximately
result in the same differences for the overall desorption temperature.
We thus estimate a decrease of desorption temperature of \MG\ when
alloying with Zn for Reaction 1 of approximately 123~K for 1/3 Zn
concentration and 262~K for 2/3 Zn. Note that we obtain almost identical
numbers using a different argument: Simply evaluating the experimental
relationship in Eq.~(\ref{equ:fit_H}), purely based on our calculated
changes in enthalpy, gives a lowering of the desorption temperature of
125~K for 1/3 Zn concentration and 250~K for 2/3 Zn.

From our results we see that the formation of the intermediate
MgB$_{12}$H$_{12}$ is mainly responsible for the kinetic barrier, and
methods to improve the kinetics of borohydrides should focus on reaction
pathways avoiding the formation of this intermediate, as suggested by
several other groups.\cite{Yan_2011:formation_process,
Severa_2010:direct_hydrogenation,
Soloveichik_2009:magnesium_borohydride} Because the kinetic barrier is
roughly constant, at least for the full hydrogen release reaction, we do
not expect alloying to have a direct impact on the height of the kinetic
barrier. But, it is really the overall energy required to start the
hydrogen desorption that is of interest---and that linearly decreases
with Zn concentration.  Note that the reaction kinetics can, at least in
principle, be accelerated by using catalysts or by controlling the
particle size of reactants.\cite{Alapati_2007:using_first} Furthermore,
it is also conceivable that the kinetics of the dehydrogenation reaction
can be influenced by the inclusion of impurities, as studied by van de
Walle's group.\cite{Hoang_2009:hydrogen-related_defects}

%%%%%%%%%%%%%%%%%%%%%%%%%%%%%%%%%%%%%%%%%%%%%%%%%%%%%%%%%%%%%%%%%%%%%%%%
\section{Conclusions}
%%%%%%%%%%%%%%%%%%%%%%%%%%%%%%%%%%%%%%%%%%%%%%%%%%%%%%%%%%%%%%%%%%%%%%%%

In summary, \emph{ab initio} calculations were performed to accurately
determine formation enthalpies for many different materials suspected in
the decomposition of \MG, to find reaction enthalpies for the most
likely reaction pathways, and to study the effect of Zn concentration on
the overall desorption reaction.  We find that the overall
thermodynamics of the desorption reaction can be optimized by alloying
\MG\ with around 33\% Zn. We estimate that in this case the temperature
for the complete decomposition reaction is lowered by 123~K. Verifying
the kinetics explicitly through \emph{ab initio} calculations is
difficult and is the subject of ongoing research. Supported by
encouraging experimental
results,\cite{Albanese_2013:theoretical_experimental,
Kalantzopoulos_2014:hydrogen_storage} we conclude that alloying \MG\
with Zn is an ideal option for fine-tuning the desorption reaction
without sacrificing too much gravimetric storage density.

%%%%%%%%%%%%%%%%%%%%%%%%%%%%%%%%%%%%%%%%%%%%%%%%%%%%%%%%%%%%%%%%%%%%%%%%
\begin{acknowledgements}
This work was supported in full by NSF Grant No.\ DMR-1145968.
\end{acknowledgements}
%%%%%%%%%%%%%%%%%%%%%%%%%%%%%%%%%%%%%%%%%%%%%%%%%%%%%%%%%%%%%%%%%%%%%%%%

%%%%%%%%%%%%%%%%%%%%%%%%%%%%%%%%%%%%%%%%%%%%%%%%%%%%%%%%%%%%%%%%%%%%%%%%
\appendix
%%%%%%%%%%%%%%%%%%%%%%%%%%%%%%%%%%%%%%%%%%%%%%%%%%%%%%%%%%%%%%%%%%%%%%%%

%%%%%%%%%%%%%%%%%%%%%%%%%%%%%%%%%%%%%%%%%%%%%%%%%%%%%%%%%%%%%%%%%%%%%%%%
\section{Standard enthalpies of formation}
%%%%%%%%%%%%%%%%%%%%%%%%%%%%%%%%%%%%%%%%%%%%%%%%%%%%%%%%%%%%%%%%%%%%%%%%

We give here standard enthalpies of formations for all structures used
throughout the main manuscript, calculated from DFT ground-state
energies with temperature and zero-point contributions included through
Eq.~(\ref{eq:enthalpy_phonon}). Units of
temperature $T$ and enthalpy $\Delta H$ are given in K and kJ/mol,
respectively.  We used experimental data for the temperature contributed
enthalpy of H$_2$ gas.\cite{Hemmes_1986:thermodynamic_properties}

%%%%%%%%%%%%%%%%%%%%%%%%%%%%%%%%%%%%%%%%%%%%%%%%%%%%%%%%%%%%%%%%%%%%%%%%
\onecolumngrid
\subsection{MgB$_2$}
%%%%%%%%%%%%%%%%%%%%%%%%%%%%%%%%%%%%%%%%%%%%%%%%%%%%%%%%%%%%%%%%%%%%%%%%

\vspace{-2ex}\noindent
\begin{tabular*}{\textwidth}{@{}>{$}l<{$}@{\extracolsep{\fill}}
*6{>{$}c<{$}}>{$}r<{$}@{}}\hline\hline
T\text{ [K]} & \Delta H\text{ [kJ/mol]}  &   T\text{ [K]} & \Delta H\text{ [kJ/mol]}&
T\text{ [K]} & \Delta H\text{ [kJ/mol]}  &   T\text{ [K]} & \Delta H\text{ [kJ/mol]}\\\hline
25   &   -42.8478   &      225  &   -43.5773   &     425  &   -43.0847   &     625  &   -42.5631\\
50   &   -42.9136   &      250  &   -43.5474   &     450  &   -43.0123   &     650  &   -42.5079\\
75   &   -43.0676   &      275  &   -43.5007   &     475  &   -42.9416   &     675  &   -42.4547\\
100  &   -43.2423   &      300  &   -43.4422   &     500  &   -42.8729   &     700  &   -42.4035\\
125  &   -43.3909   &      325  &   -43.3760   &     525  &   -42.8063   &     725  &   -42.3542\\
150  &   -43.4969   &      350  &   -43.3053   &     550  &   -42.7421   &     750  &   -42.3067\\
175  &   -43.5594   &      375  &   -43.2321   &     575  &   -42.6802   &     775  &   -42.2608\\
200  &   -43.5837   &      400  &   -43.1582   &     600  &   -42.6205   &     800  &   -42.2166\\\hline\hline
\end{tabular*}\\[1ex]
$\Delta H (T=0) = -42.8445$ kJ/mol\\
$\Delta H (E\text{ only}) = -41.3923$ kJ/mol
\vspace{-1ex}

%%%%%%%%%%%%%%%%%%%%%%%%%%%%%%%%%%%%%%%%%%%%%%%%%%%%%%%%%%%%%%%%%%%%%%%%
\subsection{MgH$_2$}
%%%%%%%%%%%%%%%%%%%%%%%%%%%%%%%%%%%%%%%%%%%%%%%%%%%%%%%%%%%%%%%%%%%%%%%%

\vspace{-2ex}\noindent
\begin{tabular*}{\textwidth}{@{}>{$}l<{$}@{\extracolsep{\fill}}
*6{>{$}c<{$}}>{$}r<{$}@{}}\hline\hline
T\text{ [K]} & \Delta H\text{ [kJ/mol]}  &   T\text{ [K]} & \Delta H\text{ [kJ/mol]}&
T\text{ [K]} & \Delta H\text{ [kJ/mol]}  &   T\text{ [K]} & \Delta H\text{ [kJ/mol]}\\\hline
25   &   -71.2620   &     225  &   -77.1642    &    425  &   -80.2792   &     625 &    -80.6338\\
50   &   -72.0536   &     250  &   -77.7319    &    450  &   -80.4525   &     650 &    -80.5411\\
75   &   -72.8916   &     275  &   -78.2442    &    475  &   -80.5836   &     675 &    -80.4248\\
100  &   -73.7396   &     300  &   -78.7003    &    500  &   -80.6754   &     700 &    -80.2867\\
125  &   -74.4942   &     325  &   -79.1201    &    525  &   -80.7307   &     725 &    -80.1277\\
150  &   -75.2186   &     350  &   -79.4855    &    550  &   -80.7519   &     750 &    -79.9499\\
175  &   -75.9036   &     375  &   -79.7983    &    575  &   -80.7413   &     775 &    -79.7546\\
200  &   -76.5428   &     400  &   -80.0609    &    600  &   -80.7012   &     800 &    -79.5430\\ \hline\hline
\end{tabular*}\\[1ex]
$\Delta H (T=0) = -70.5096$ kJ/mol\\
$\Delta H (E\text{ only}) = -80.7888$ kJ/mol
\vspace{-1ex}

%%%%%%%%%%%%%%%%%%%%%%%%%%%%%%%%%%%%%%%%%%%%%%%%%%%%%%%%%%%%%%%%%%%%%%%%
\subsection{MgB$_{12}$H$_{12}$}
%%%%%%%%%%%%%%%%%%%%%%%%%%%%%%%%%%%%%%%%%%%%%%%%%%%%%%%%%%%%%%%%%%%%%%%%

\vspace{-2ex}\noindent
\begin{tabular*}{\textwidth}{@{}>{$}l<{$}@{\extracolsep{\fill}}
*6{>{$}c<{$}}>{$}r<{$}@{}}\hline\hline
T\text{ [K]} & \Delta H\text{ [kJ/mol]}  &   T\text{ [K]} & \Delta H\text{ [kJ/mol]}&
T\text{ [K]} & \Delta H\text{ [kJ/mol]}  &   T\text{ [K]} & \Delta H\text{ [kJ/mol]}\\\hline
25   &   -474.2918   &    225  &   -503.3760   &    425  &   -523.9349   &    625  &   -532.7084\\
50   &   -478.3169   &    250  &   -506.6553   &    450  &   -525.6071   &    650  &   -533.1555\\
75   &   -482.1853   &    275  &   -509.7199   &    475  &   -527.0971   &    675  &   -533.4792\\
100  &   -486.0506   &    300  &   -512.5552   &    500  &   -528.4145   &    700  &   -533.6845\\
125  &   -489.5086   &    325  &   -515.2722   &    525  &   -529.5698   &    725  &   -533.7732\\
150  &   -493.0109   &    350  &   -517.7585   &    550  &   -530.5696   &    750  &   -533.7528\\
175  &   -496.5009   &    375  &   -520.0212   &    575  &   -531.4215   &    775  &   -533.6277\\
200  &   -499.9106   &    400  &   -522.0699   &    600  &   -532.1323   &    800  &   -533.4021\\ \hline\hline
\end{tabular*}\\[1ex]
$\Delta H (T=0) = -469.9027$ kJ/mol\\
$\Delta H (E\text{ only}) = -612.5082$ kJ/mol
\vspace{-1ex}

%%%%%%%%%%%%%%%%%%%%%%%%%%%%%%%%%%%%%%%%%%%%%%%%%%%%%%%%%%%%%%%%%%%%%%%%
\subsection{MgB$_4$}
%%%%%%%%%%%%%%%%%%%%%%%%%%%%%%%%%%%%%%%%%%%%%%%%%%%%%%%%%%%%%%%%%%%%%%%%

\vspace{-2ex}\noindent
\begin{tabular*}{\textwidth}{@{}>{$}l<{$}@{\extracolsep{\fill}}
*6{>{$}c<{$}}>{$}r<{$}@{}}\hline\hline
T\text{ [K]} & \Delta H\text{ [kJ/mol]}  &   T\text{ [K]} & \Delta H\text{ [kJ/mol]}&
T\text{ [K]} & \Delta H\text{ [kJ/mol]}  &   T\text{ [K]} & \Delta H\text{ [kJ/mol]}\\\hline
25   &   -53.5246   &     225  &   -53.8921   &     425  &   -53.8095   &     625  &   -53.6442\\
50   &   -53.5565   &     250  &   -53.8964   &     450  &   -53.7899   &     650  &   -53.6232\\
75   &   -53.6239   &     275  &   -53.8942   &     475  &   -53.7698   &     675  &   -53.6023\\
100  &   -53.6993   &     300  &   -53.8869   &     500  &   -53.7492   &     700  &   -53.5815\\
125  &   -53.7661   &     325  &   -53.8758   &     525  &   -53.7284   &     725  &   -53.5607\\
150  &   -53.8183   &     350  &   -53.8619   &     550  &   -53.7074   &     750  &   -53.5401\\
175  &   -53.8554   &     375  &   -53.8458   &     575  &   -53.6864   &     775  &   -53.5196\\
200  &   -53.8792   &     400  &   -53.8282   &     600  &   -53.6653   &     800  &   -53.4992\\ \hline\hline
\end{tabular*}\\[1ex]
$\Delta H (T=0) = -53.5227$ kJ/mol\\
$\Delta H (E\text{ only}) = -53.8811$ kJ/mol

%%%%%%%%%%%%%%%%%%%%%%%%%%%%%%%%%%%%%%%%%%%%%%%%%%%%%%%%%%%%%%%%%%%%%%%%
\subsection{Zn-alloyed Mg(BH$_4$)$_2$}
%%%%%%%%%%%%%%%%%%%%%%%%%%%%%%%%%%%%%%%%%%%%%%%%%%%%%%%%%%%%%%%%%%%%%%%%

Values are given for all 10 alloying levels of \MG\ from 0/9 to 9/9.
Units for temperature $T$ and enthalpies $\Delta H$ are K and kJ/mol,
respectively.
\bigskip

\noindent
\begingroup\small
\begin{tabular*}{\textwidth}{@{}>{$}l<{$}@{\extracolsep{\fill}}
*9{>{$}c<{$}}>{$}r<{$}@{}}\hline\hline
T    &\Delta H (0/9)& \Delta H (1/9)& \Delta H (2/9)& \Delta H (3/9) &\Delta H (4/9)&\Delta H (5/9) &\Delta H (6/9)&\Delta H (7/9) &\Delta H (8/9) &\Delta H (9/9)\\\hline
E_{\textrm{only}}$$  & -261.7976 & -236.4970   &    -210.9513   &    -185.1490    &  -159.9197    &   -133.7350    &   -108.8261   &    -83.6754    &    -58.5632    &    -34.6722\\
0    &   -181.1409   &    -156.6974   &    -131.9971   &    -107.1175    &   -82.8319    &    -57.5765    &    -33.5969   &     -9.4840    &     14.9718    &     37.8973\\
25   &   -184.1121   &    -159.6661   &    -134.9744   &    -110.0966    &   -85.8174    &    -60.5598    &    -36.5847   &     -12.4716   &     11.9706    &     34.8856\\
50   &   -186.9262   &    -162.4711   &    -137.7886   &    -112.9080    &   -88.6331    &    -63.3640    &    -39.3886   &     -15.2746   &     9.1561     &     32.0606\\
75   &   -189.5518   &    -165.0799   &    -140.3934   &    -115.4973    &   -91.2175    &    -65.9289    &    -41.9403   &     -17.8210   &     6.6122     &     29.5198\\
100  &   -191.9842   &    -167.4923   &    -142.7930   &    -117.8747    &   -93.5821    &    -68.2694    &    -44.2600   &     -20.1315   &     4.3123     &     27.2310\\
125  &   -193.9260   &    -169.4133   &    -144.6961   &    -119.7533    &   -95.4431    &    -70.1046    &    -46.0719   &     -21.9319   &     2.5255     &     25.4591\\
150  &   -195.7116   &    -171.1788   &    -146.4412   &    -121.4734    &   -97.1444    &    -71.7801    &    -47.7249   &     -23.5728   &     0.8988     &     23.8480\\
175  &   -197.3677   &    -172.8158   &    -148.0571   &    -123.0648    &   -98.7177    &    -73.3288    &    -49.2532   &     -25.0893   &     -0.6041    &     22.3604\\
200  &   -198.9169   &    -174.3473   &    -149.5675   &    -124.5518    &   -100.1878   &    -74.7759    &    -50.6822   &     -26.5069   &     -2.0090    &     20.9701\\
225  &   -200.5059   &    -175.9197   &    -151.1194   &    -126.0814    &   -101.7016   &    -76.2683    &    -52.1583   &     -27.9718   &     -3.4620    &     19.5312\\
250  &   -202.0122   &    -177.4102   &    -152.5902   &    -127.5308    &   -103.1361   &    -77.6828    &    -53.5577   &     -29.3599   &     -4.8388    &     18.1682\\
275  &   -203.4391   &    -178.8220   &    -153.9828   &    -128.9030    &   -104.4940   &    -79.0218    &    -54.8824   &     -30.6729   &     -6.1412    &     16.8799\\
300  &   -204.7866   &    -180.1550   &    -155.2971   &    -130.1978    &   -105.7747   &    -80.2844    &    -56.1313   &     -31.9097   &     -7.3677    &     15.6677\\
325  &   -206.1304   &    -181.4847   &    -156.6087   &    -131.4905    &   -107.0534   &    -81.5459    &    -57.3793   &     -33.1450   &     -8.5931    &     14.4568\\
350  &   -207.3915   &    -182.7322   &    -157.8386   &    -132.7021    &   -108.2511   &    -82.7269    &    -58.5470   &     -34.2996   &     -9.7381    &     13.3268\\
375  &   -208.5682   &    -183.8955   &    -158.9848   &    -133.8306    &   -109.3658   &    -83.8256    &    -59.6324   &     -35.3715   &     -10.8005   &     12.2795\\
400  &   -209.6592   &    -184.9734   &    -160.0461   &    -134.8747    &   -110.3961   &    -84.8403    &    -60.6339   &     -36.3591   &     -11.7789   &     11.3164\\
425  &   -210.6765   &    -185.9779   &    -161.0344   &    -135.8465    &   -111.3540   &    -85.7832    &    -61.5637   &     -37.2747   &     -12.6855   &     10.4253\\
450  &   -211.6066   &    -186.8955   &    -161.9364   &    -136.7324    &   -112.2263   &    -86.6409    &    -62.4083   &     -38.1048   &     -13.5068   &     9.6196 \\
475  &   -212.4497   &    -187.7263   &    -162.7521   &    -137.5326    &   -113.0128   &    -87.4133    &    -63.1678   &     -38.8497   &     -14.2430   &     8.8989 \\
500  &   -213.2061   &    -188.4708   &    -163.4819   &    -138.2475    &   -113.7141   &    -88.1010    &    -63.8425   &     -39.5098   &     -14.8946   &     8.2629 \\
525  &   -213.8777   &    -189.1306   &    -164.1276   &    -138.8788    &   -114.3320   &    -88.7056    &    -64.4344   &     -40.0870   &     -15.4635   &     7.7095 \\
550  &   -214.4644   &    -189.7059   &    -164.6892   &    -139.4265    &   -114.8665   &    -89.2273    &    -64.9435   &     -40.5814   &     -15.9497   &     7.2385 \\
575  &   -214.9676   &    -190.1980   &    -165.1681   &    -139.8920    &   -115.3189   &    -89.6673    &    -65.3711   &     -40.9945   &     -16.3547   &     6.8486 \\
600  &   -215.3889   &    -190.6084   &    -165.5658   &    -140.2768    &   -115.6908   &    -90.0273    &    -65.7189   &     -41.3278   &     -16.6801   &     6.5381 \\
625  &   -215.7300   &    -190.9390   &    -165.8841   &    -140.5827    &   -115.9840   &    -90.3089    &    -65.9885   &     -41.5831   &     -16.9276   &     6.3052 \\
650  &   -215.9928   &    -191.1914   &    -166.1248   &    -140.8114    &   -116.2003   &    -90.5139    &    -66.1818   &     -41.7623   &     -17.0992   &     6.1480 \\
675  &   -216.1793   &    -191.3679   &    -166.2899   &    -140.9650    &   -116.3418   &    -90.6445    &    -66.3009   &     -41.8675   &     -17.1969   &     6.0644 \\
700  &   -216.2917   &    -191.4705   &    -166.3815   &    -141.0455    &   -116.4104   &    -90.7026    &    -66.3477   &     -41.9007   &     -17.2227   &     6.0522 \\
725  &   -216.3300   &    -191.4993   &    -166.3998   &    -141.0531    &   -116.4063   &    -90.6884    &    -66.3225   &     -41.8621   &     -17.1769   &     6.1115 \\
750  &   -216.2986   &    -191.4587   &    -166.3489   &    -140.9920    &   -116.3338   &    -90.6061    &    -66.2294   &     -41.7559   &     -17.0636   &     6.2378 \\
775  &   -216.1996   &    -191.3507   &    -166.2312   &    -140.8643    &   -116.1951   &    -90.4578    &    -66.0707   &     -41.5843   &     -16.8850   &     6.4290 \\
800  &   -216.0355   &    -191.1778   &    -166.0488   &    -140.6724    &   -115.9924   &    -90.2459    &    -65.8486   &     -41.3496   &     -16.6435   &     6.6828 \\ \hline\hline
\end{tabular*}\\[1ex]
\endgroup

%%%%%%%%%%%%%%%%%%%%%%%%%%%%%%%%%%%%%%%%%%%%%%%%%%%%%%%%%%%%%%%%%%%%%%%%
\newpage\twocolumngrid
\section{Absolute entropies}
%%%%%%%%%%%%%%%%%%%%%%%%%%%%%%%%%%%%%%%%%%%%%%%%%%%%%%%%%%%%%%%%%%%%%%%%

We give here absolute entropies for all structures used throughout the
main manuscript, calculated using Eq.~(\ref{eq:entropy_phonon}).
Units for temperature $T$ and entropies $S$ are K and J/K/mol,
respectively.

%%%%%%%%%%%%%%%%%%%%%%%%%%%%%%%%%%%%%%%%%%%%%%%%%%%%%%%%%%%%%%%%%%%%%%%%
\onecolumngrid
\subsection{MgB$_2$}
%%%%%%%%%%%%%%%%%%%%%%%%%%%%%%%%%%%%%%%%%%%%%%%%%%%%%%%%%%%%%%%%%%%%%%%%

\noindent
\begin{tabular*}{\textwidth}{@{}>{$}l<{$}@{\extracolsep{\fill}}
*6{>{$}c<{$}}>{$}r<{$}@{}}\hline\hline
T\text{ [K]} & S\text{ [J/K/mol]}  &   T\text{ [K]} & S\text{ [J/K/mol]}&
T\text{ [K]} & S\text{ [J/K/mol]}  &   T\text{ [K]} & S\text{ [J/K/mol]}\\\hline
25   &   0.0049   &      225  &   23.2048   &     425  &   55.1436   &     625  &   79.7405\\
50   &   0.2642   &      250  &   27.5212   &     450  &   58.6030   &     650  &   82.3799\\
75   &   1.4697   &      275  &   31.7893   &     475  &   61.9427   &     675  &   84.9384\\
100  &   3.7830   &      300  &   35.9743   &     500  &   65.1672   &     700  &   87.4202\\
125  &   6.9282   &      325  &   40.0539   &     525  &   68.2817   &     725  &   89.8290\\
150  &   10.6248  &      350  &   44.0151   &     550  &   71.2912   &     750  &   92.1687\\
175  &   14.6636  &      375  &   47.8514   &     575  &   74.2008   &     775  &   94.4427\\
200  &   18.8933  &      400  &   51.5607   &     600  &   77.0157   &     800  &   96.6542\\ \hline\hline
\end{tabular*}

%%%%%%%%%%%%%%%%%%%%%%%%%%%%%%%%%%%%%%%%%%%%%%%%%%%%%%%%%%%%%%%%%%%%%%%%
\subsection{MgH$_2$}
%%%%%%%%%%%%%%%%%%%%%%%%%%%%%%%%%%%%%%%%%%%%%%%%%%%%%%%%%%%%%%%%%%%%%%%%

\noindent
\begin{tabular*}{\textwidth}{@{}>{$}l<{$}@{\extracolsep{\fill}}
*6{>{$}c<{$}}>{$}r<{$}@{}}\hline\hline
T\text{ [K]} & S\text{ [J/K/mol]}  &   T\text{ [K]} & S\text{ [J/K/mol]}&
T\text{ [K]} & S\text{ [J/K/mol]}  &   T\text{ [K]} & S\text{ [J/K/mol]}\\\hline
25   &   0.0301   &     225  &   21.1051    &    425  &   43.7259   &     625 &    63.4665\\
50   &   0.8668   &     250  &   24.0625    &    450  &   46.3667   &     650 &    65.7097\\
75   &   2.9991   &     275  &   26.9837    &    475  &   48.9595   &     675 &    67.9047\\
100  &   5.8224   &     300  &   29.8698    &    500  &   51.5030   &     700 &    70.0524\\
125  &   8.8840   &     325  &   32.7202    &    525  &   53.9965   &     725 &    72.1538\\
150  &   11.9888  &     350  &   35.5334    &    550  &   56.4395   &     750 &    74.2101\\
175  &   15.0695  &     375  &   38.3071    &    575  &   58.8319   &     775 &    76.2224\\
200  &   18.1085  &     400  &   41.0387    &    600  &   61.1741   &     800 &    78.1921\\ \hline\hline
\end{tabular*}

%%%%%%%%%%%%%%%%%%%%%%%%%%%%%%%%%%%%%%%%%%%%%%%%%%%%%%%%%%%%%%%%%%%%%%%%
\subsection{MgB$_{12}$H$_{12}$}
%%%%%%%%%%%%%%%%%%%%%%%%%%%%%%%%%%%%%%%%%%%%%%%%%%%%%%%%%%%%%%%%%%%%%%%%

\noindent
\begin{tabular*}{\textwidth}{@{}>{$}l<{$}@{\extracolsep{\fill}}
*6{>{$}c<{$}}>{$}r<{$}@{}}\hline\hline
T\text{ [K]} & S\text{ [J/K/mol]}  &   T\text{ [K]} & S\text{ [J/K/mol]}&
T\text{ [K]} & S\text{ [J/K/mol]}  &   T\text{ [K]} & S\text{ [J/K/mol]}\\\hline
25   &   6.3137   &    225  &   120.6799   &    425  &   270.7316   &    625  &   415.3139\\
50   &   20.9873  &    250  &   137.9110   &    450  &   289.7719   &    650  &   431.9936\\
75   &   35.4722  &    275  &   155.9381   &    475  &   308.5848   &    675  &   448.3569\\
100  &   48.9467  &    300  &   174.5592   &    500  &   327.1345   &    700  &   464.4084\\
125  &   61.9336  &    325  &   193.5822   &    525  &   345.3956   &    725  &   480.1542\\
150  &   75.1936  &    350  &   212.8384   &    550  &   363.3518   &    750  &   495.6008\\
175  &   89.2742  &    375  &   232.1869   &    575  &   380.9930   &    775  &   510.7553\\
200  &   104.4286 &    400  &   251.5141   &    600  &   398.3141   &    800  &   525.6249\\ \hline\hline
\end{tabular*}

%%%%%%%%%%%%%%%%%%%%%%%%%%%%%%%%%%%%%%%%%%%%%%%%%%%%%%%%%%%%%%%%%%%%%%%%
\subsection{MgB$_4$}
%%%%%%%%%%%%%%%%%%%%%%%%%%%%%%%%%%%%%%%%%%%%%%%%%%%%%%%%%%%%%%%%%%%%%%%%

\noindent
\begin{tabular*}{\textwidth}{@{}>{$}l<{$}@{\extracolsep{\fill}}
*6{>{$}c<{$}}>{$}r<{$}@{}}\hline\hline
T\text{ [K]} & S\text{ [J/K/mol]}  &   T\text{ [K]} & S\text{ [J/K/mol]}&
T\text{ [K]} & S\text{ [J/K/mol]}  &   T\text{ [K]} & S\text{ [J/K/mol]}\\\hline
25   &   0.0637   &     225  &   34.4447   &     425  &   80.6957   &     625  &   118.8385\\
50   &   1.1818   &     250  &   40.4270   &     450  &   85.9533   &     650  &   123.0205\\
75   &   3.8995   &     275  &   46.4271   &     475  &   91.0663   &     675  &   127.0873\\
100  &   7.7271   &     300  &   52.3932   &     500  &   96.0356   &     700  &   131.0436\\
125  &   12.2816  &     325  &   58.2867   &     525  &   100.8636  &     725  &   134.8940\\
150  &   17.3581  &     350  &   64.0798   &     550  &   105.5537  &     750  &   138.6430\\
175  &   22.8185  &     375  &   69.7532   &     575  &   110.1098  &     775  &   142.2949\\
200  &   28.5473  &     400  &   75.2944   &     600  &   114.5366  &     800  &   145.8538\\ \hline\hline
\end{tabular*}

%%%%%%%%%%%%%%%%%%%%%%%%%%%%%%%%%%%%%%%%%%%%%%%%%%%%%%%%%%%%%%%%%%%%%%%%
\subsection{Zn-alloyed Mg(BH$_4$)$_2$}
%%%%%%%%%%%%%%%%%%%%%%%%%%%%%%%%%%%%%%%%%%%%%%%%%%%%%%%%%%%%%%%%%%%%%%%%

Values are given for all 10 alloying levels of \MG\ from 0/9 to 9/9.
Units for temperature $T$ and entropy $S$ are K and J/K/mol,
respectively.

\noindent
\begingroup\small
\begin{tabular*}{\textwidth}{@{}>{$}l<{$}@{\extracolsep{\fill}}
*9{>{$}c<{$}}>{$}r<{$}@{}}\hline\hline
T    &   S (0/9)    &        S (1/9) &         S (2/9)& S (3/9)        &       S (4/9)   &       S (5/9) &      S (6/9)   &        S (7/9) &        S (8/9) &      S (9/9)\\ \hline
25   &   1.6220     &      2.2841    &       2.2170   &       2.6043   &       2.7088    &      3.3761   &       3.5663   &       4.1165   &       3.6454   &       3.4832\\
50   &   8.3785     &      9.8673    &      10.1158   &      11.1473   &      11.7055    &     13.2631   &      14.0306   &      15.1895   &      14.9646   &      15.0737\\
75   &   18.0017    &      20.0819   &      20.7124   &      22.3144   &      23.2698    &     25.4639   &      26.7622   &      28.3285   &      28.4614   &      28.9387\\
100  &   28.7955    &      31.2783   &      32.2271   &      34.2544   &      35.5285    &     38.1711   &      39.8800   &      41.7232   &      42.1489   &      42.9248\\
125  &   39.9059    &      42.6715   &      43.8770   &      46.2207   &      47.7481    &     50.7195   &      52.7340   &      54.7764   &      55.4214   &      56.4269\\
150  &   50.9458    &      53.9177   &      55.3307   &      57.9162   &      59.6395    &     62.8582   &      65.0962   &      67.2857   &      68.0931   &      69.2724\\
175  &   61.7268    &      64.8542   &      66.4356   &      69.2103   &      71.0836    &     74.4918   &      76.8938   &      79.1943   &      80.1235   &      81.4350\\
200  &   72.1649    &      75.4130   &      77.1325   &      80.0582   &      82.0477    &     85.6044   &      88.1291   &      90.5164   &      91.5394   &      92.9545\\
225  &   82.2382    &      85.5830   &      87.4171   &      90.4661   &      92.5482    &     96.2238   &      98.8438   &     101.3023   &     102.3994   &     103.8990\\
250  &   91.9590    &      95.3833   &      97.3144   &     100.4663   &     102.6246    &    106.3980   &     109.0946   &     111.6142   &     112.7719   &     114.3435\\
275  &   101.3544   &     104.8461   &     106.8603   &     110.0999   &     112.3230    &    116.1785   &     118.9395   &     121.5135   &     122.7219   &     124.3569\\
300  &   110.4554   &     114.0053   &     116.0920   &     119.4075   &     121.6873    &    125.6133   &     128.4297   &     131.0536   &     132.3055   &     133.9978\\
325  &   119.2913   &     122.8923   &     125.0430   &     128.4251   &     130.7555    &    134.7429   &     137.6085   &     140.2788   &     141.5685   &     143.3135\\
350  &   127.8875   &     131.5339   &     133.7417   &     137.1828   &     139.5591    &    143.6006   &     146.5106   &     149.2246   &     150.5477   &     152.3417\\
375  &   136.2653   &     139.9525   &     142.2114   &     145.7053   &     148.1238    &    152.2135   &     155.1640   &     157.9193   &     159.2723   &     161.1119\\
400  &   144.4423   &     148.1666   &     150.4716   &     154.0130   &     156.4703    &    160.6034   &     163.5911   &     166.3856   &     167.7655   &     169.6480\\
425  &   152.4333   &     156.1913   &     158.5381   &     162.1224   &     164.6158    &    168.7881   &     171.8103   &     174.6419   &     176.0464   &     177.9691\\
450  &   160.2505   &     164.0394   &     166.4242   &     170.0474   &     172.5745    &    176.7823   &     179.8367   &     182.7035   &     184.1304   &     186.0909\\
475  &   167.9043   &     171.7216   &     174.1411   &     177.7996   &     180.3582    &    184.5984   &     187.6828   &     190.5831   &     192.0306   &     194.0267\\
500  &   175.4037   &     179.2471   &     181.6983   &     185.3892   &     187.9771    &    192.2471   &     195.3595   &     198.2915   &     199.7581   &     201.7877\\
525  &   182.7562   &     186.6239   &     189.1041   &     192.8245   &     195.4401    &    199.7373   &     202.8760   &     205.8380   &     207.3223   &     209.3835\\
550  &   189.9688   &     193.8589   &     196.3658   &     200.1133   &     202.7547    &    207.0769   &     210.2403   &     213.2308   &     214.7316   &     216.8225\\
575  &   197.0471   &     200.9582   &     203.4895   &     207.2619   &     209.9277    &    214.2730   &     217.4595   &     220.4770   &     221.9932   &     224.1120\\
600  &   203.9967   &     207.9271   &     210.4811   &     214.2763   &     216.9650    &    221.3316   &     224.5399   &     227.5829   &     229.1135   &     231.2586\\
625  &   210.8220   &     214.7706   &     217.3453   &     221.1617   &     223.8719    &    228.2583   &     231.4870   &     234.5542   &     236.0983   &     238.2681\\
650  &   217.5274   &     221.4929   &     224.0869   &     227.9228   &     230.6532    &    235.0579   &     238.3058   &     241.3958   &     242.9527   &     245.1457\\
675  &   224.1166   &     228.0979   &     230.7098   &     234.5638   &     237.3132    &    241.7350   &     245.0009   &     248.1126   &     249.6814   &     251.8963\\
700  &   230.5933   &     234.5894   &     237.2179   &     241.0885   &     243.8558    &    248.2935   &     251.5764   &     254.7085   &     256.2887   &     258.5241\\
725  &   236.9606   &     240.9707   &     243.6144   &     247.5006   &     250.2848    &    254.7373   &     258.0361   &     261.1876   &     262.7785   &     265.0332\\
750  &   243.2216   &     247.2447   &     249.9027   &     253.8033   &     256.6034    &    261.0697   &     264.3836   &     267.5534   &     269.1543   &     271.4272\\
775  &   249.3791   &     253.4144   &     256.0858   &     259.9998   &     262.8148    &    267.2940   &     270.6221   &     273.8092   &     275.4197   &     277.7096\\
800  &   255.4358   &     259.4825   &     262.1663   &     266.0928   &     268.9220    &    273.4133   &     276.7547   &     279.9581   &     281.5777   &     283.8837\\ \hline\hline
\end{tabular*}

%%%%%%%%%%%%%%%%%%%%%%%%%%%%%%%%%%%%%%%%%%%%%%%%%%%%%%%%%%%%%%%%%%%%%%%%
\subsection{Mg}
%%%%%%%%%%%%%%%%%%%%%%%%%%%%%%%%%%%%%%%%%%%%%%%%%%%%%%%%%%%%%%%%%%%%%%%%

\noindent
\begin{tabular*}{\textwidth}{@{}>{$}l<{$}@{\extracolsep{\fill}}
*6{>{$}c<{$}}>{$}r<{$}@{}}\hline\hline
T\text{ [K]} & S\text{ [J/K/mol]}  &   T\text{ [K]} & S\text{ [J/K/mol]}&
T\text{ [K]} & S\text{ [J/K/mol]}  &   T\text{ [K]} & S\text{ [J/K/mol]}\\\hline
25   &   0.1591   &     225  &   25.4974   &     425  &   40.3539  &     625  &   49.6940\\
50   &   2.0640   &     250  &   27.8794   &     450  &   41.7287  &     650  &   50.6506\\
75   &   5.6107   &     275  &   30.0685   &     475  &   43.0327  &     675  &   51.5719\\
100  &   9.5667   &     300  &   32.0909   &     500  &   44.2727  &     700  &   52.4604\\
125  &   13.3697  &     325  &   33.9685   &     525  &   45.4546  &     725  &   53.3183\\
150  &   16.8606  &     350  &   35.7196   &     550  &   46.5835  &     750  &   54.1477\\
175  &   20.0248  &     375  &   37.3593   &     575  &   47.6638  &     775  &   54.9504\\
200  &   22.8912  &     400  &   38.9005   &     600  &   48.6995  &     800  &   55.7281\\ \hline\hline
\end{tabular*}

%%%%%%%%%%%%%%%%%%%%%%%%%%%%%%%%%%%%%%%%%%%%%%%%%%%%%%%%%%%%%%%%%%%%%%%%
\subsection{B}
%%%%%%%%%%%%%%%%%%%%%%%%%%%%%%%%%%%%%%%%%%%%%%%%%%%%%%%%%%%%%%%%%%%%%%%%

\noindent
\begin{tabular*}{\textwidth}{@{}>{$}l<{$}@{\extracolsep{\fill}}
*6{>{$}c<{$}}>{$}r<{$}@{}}\hline\hline
T\text{ [K]} & S\text{ [J/K/mol]}  &   T\text{ [K]} & S\text{ [J/K/mol]}&
T\text{ [K]} & S\text{ [J/K/mol]}  &   T\text{ [K]} & S\text{ [J/K/mol]}\\\hline
25   &   0.0000  &     225  &   3.3007   &     425  &   11.0926  &     625  &   18.2138\\
50   &   0.0061  &     250  &   4.2054   &     450  &   12.0522  &     650  &   19.0120\\
75   &   0.0690  &     275  &   5.1561   &     475  &   12.9935  &     675  &   19.7904\\
100  &   0.2541  &     300  &   6.1357   &     500  &   13.9153  &     700  &   20.5498\\
125  &   0.5915  &     325  &   7.1308   &     525  &   14.8167  &     725  &   21.2906\\
150  &   1.0835  &     350  &   8.1310   &     550  &   15.6972  &     750  &   22.0135\\
175  &   1.7150  &     375  &   9.1284   &     575  &   16.5568  &     775  &   22.7191\\
200  &   2.4626  &     400  &   10.117   &     600  &   17.3956  &     800  &   23.4080\\ \hline\hline
\end{tabular*}

%%%%%%%%%%%%%%%%%%%%%%%%%%%%%%%%%%%%%%%%%%%%%%%%%%%%%%%%%%%%%%%%%%%%%%%%
\subsection{Zn}
%%%%%%%%%%%%%%%%%%%%%%%%%%%%%%%%%%%%%%%%%%%%%%%%%%%%%%%%%%%%%%%%%%%%%%%%

\noindent
\begin{tabular*}{\textwidth}{@{}>{$}l<{$}@{\extracolsep{\fill}}
*6{>{$}c<{$}}>{$}r<{$}@{}}\hline\hline
T\text{ [K]} & S\text{ [J/K/mol]}  &   T\text{ [K]} & S\text{ [J/K/mol]}&
T\text{ [K]} & S\text{ [J/K/mol]}  &   T\text{ [K]} & S\text{ [J/K/mol]}\\\hline
25   &   4.7968   &     225  &   42.0395   &     425  &   57.3580  &     625  &   66.7984\\
50   &   12.0016  &     250  &   44.5407   &     450  &   58.7529  &     650  &   67.7615\\
75   &   18.4567  &     275  &   46.8192   &     475  &   60.0739  &     675  &   68.6886\\
100  &   23.9634  &     300  &   48.9104   &     500  &   61.3285  &     700  &   69.5823\\
125  &   28.6435  &     325  &   50.8420   &     525  &   62.5229  &     725  &   70.4450\\
150  &   32.6659  &     350  &   52.6362   &     550  &   63.6627  &     750  &   71.2786\\
175  &   36.1722  &     375  &   54.3109   &     575  &   64.7525  &     775  &   72.0851\\
200  &   39.2701  &     400  &   55.8807   &     600  &   65.7965  &     800  &   72.8662\\ \hline\hline
\end{tabular*}

%%%%%%%%%%%%%%%%%%%%%%%%%%%%%%%%%%%%%%%%%%%%%%%%%%%%%%%%%%%%%%%%%%%%%%%%
\bigskip\bigskip\bigskip\twocolumngrid

%%%%%%%%%%%%%%%%%%%%%%%%%%%%%%%%%%%%%%%%%%%%%%%%%%%%%%%%%%%%%%%%%%%%%%%%

%%%%%%%%%%%%%%%%%%%%%%%%%%%%%%%%%%%%%%%%%%%%%%%%%%%%%%%%%%%%%%%%%%%%%%%%
\end{document}